\documentclass[12pt,a4paper,final]{iopart}
\usepackage{iopams}  
\usepackage{graphicx}
\usepackage{dcolumn}  
\usepackage{bm}       
\usepackage{amssymb} 
\usepackage{braket}
\usepackage{mathcomp}

\usepackage{bbold}
\expandafter\let\csname equation*\endcsname\relax
\expandafter\let\csname endequation*\endcsname\relax 
\usepackage{amsmath}
\usepackage{physics}

\begin{document}

\title[]{The Born rule in a timeless Universe}
\vspace{3pc}

\author{Ovidiu Racorean}
\address{General Direction of Information Technology}
\address{Banul Antonache str. 52-60, sc.C, ap.19, Bucharest, Romania}
\ead{ovidiu.racorean@mfinante.gov.ro}
\vspace{3pc}

\begin{abstract}
\vspace{1pc}

In canonical quantization of gravity the wave function of the universe is CPT invariant. Thus, if the quantum state of the universe contains a particular history, than it must contain, with the same probability, the time-reversed image of that history as well. In this work, we investigate the meaning of this statement in the context of the conditional probability interpretation of Page and Wootters. Accordingly, we show that a time-reversed history of the universe is consistent with the Page and Wootters mechanism and we derive a time-reversed Schrodinger equation for the evolution of the rest of the universe. Since the same particular quantum state is acquired both in an individual history with time running forward and in its time-reversed image history, we demonstrate that conditioning the state of the universe on the hands of the clock showing a specific time results in the probability of finding the rest of the universe system in a particular state. In this scenario, we conjecture that probability interpretation is an emergent property of the quantum world.

\end{abstract}



\maketitle

\section{Introduction}

In a paper \cite{haw} published in 1985, Hawking claimed, influenced by  the CPT invariance of the wave function, that if the universe would recollapse after the moment of maximum expansion, than the arrow of time would reverse to agree with the cosmological arrow.  In other words, the universe would experience a time-reversed history, in the contracting phase. This way, an individual history of the universe would experience a time-symmetric evolution with a forward arrow of time in the expansion phase and a backward arrow of time in the contracting phase. 

Not long after, Hawking himself considered this idea as his  ”greatest mistake” when Page published a paper correcting him. Page argued in \cite{page} that not all individual histories contained in the quantum state would experience reversal of time in the contracting phase. Accordingly, there are histories of the universe in which the time runs forward during both the expansion and contraction, and there are histories with equal probability in which time runs backward in the contracting phase.

Later, Hawking has completely excluded the time-symmetric individual histories \cite{laf}such that the quantum state of the universe could contain only histories in which the arrow of time would not reverse in the contracting phase. Although individual histories would not experience a reversed arrow of time after the moment of the maximum expansion, the CPT invariance of the wave function of the universe ensures the presence of the time-reverse histories in the quantum states. Accordingly, if the quantum state of the universe contains an individual history than it must also contain with the same probability its time-reversed counterpart. Thus, this statement can be distilled to the simple conclusion that the quantum state of the universe must contain individual history and with the same probability their time-reversed counterparts.

In this paper we investigate the meaning of this conclusion in the context of Page and Wootters (PaW) conditional probability interpretation \cite{pag}, \cite{woo} of a timeless universe. In the PaW formalism time is defined operationally as a measurement realized on a subsystem of the universe serving as a clock. The rest of the universe is seen as another subsystem that is entangled with the clock. Thus, the time evolution of the rest of the universe subsystem can be tracked by conditioning the quantum state of the universe on different measurements of time on the clock subsystem. The conditional state of the rest of the universe satisfies the Schrodinger equation. In this way, we can emphasize here that an individual history of the rest of the universe consists in a series of the states of the universe conditioned on time measured in an increased sequence on the clock subsystem. 

Now, we recall that the quantum state of the universe must contain with the same probability the time-reversed histories. In this scenario we consider individual histories of the rest of the universe as sequences of conditional states of the universe in reversed order of time. In doing so, we demonstrate that conditional state of the rest of the universe satisfies a reversed-time Schrodinger equation. Thus, we advocate that time-reversal of individual histories contained in the quantum state of the universe are consistent with the conditional probability interpretation. 

Since the quantum state of the universe contains with the same probability both an individual history and its time-reversal counterpart when the state of the universe is conditioned on time as measured on the clock system, one finds two states of the rest of the universe system corresponding to the opposed arrows of time individual histories. Accordingly, the conditional state of the universe on a certain time measured on the clock subsystem takes the form of the Born rule. 

We conclude that the presence with the same probability of opposed arrows of time individual histories in the quantum state is the reason for the probabilistic nature of the quantum world, i.e., the Born rule.

\section{ The arrow of time in quantum cosmology}

In 1958, at the 11th Solvay Conference on Physics with the theme “The structure and evolution of the universe”, Gold \cite{gold} has presented the idea that the thermodynamic arrow of time and the cosmological arrow of time should always point in the same direction. This suggestion of the alignment of the arrows of time had a great impression on Hawking. Thus, under this impression and encouraged by the CPT invariance of the wave function of the universe Hawking had argued in \cite{haw} that a recollapsing universe should reverse the arrow of time after the moment of maximum expansion in order to agree with the cosmological arrow.  

Hawking’s idea of time-symmetric histories contained in the quantum state of the universe proved to be incorrect by Page in \cite{page}. Thus, Page demonstrated that not all histories contained in the quantum state had to be time-symmetric. Accordingly, there are histories of the universe in which the time runs forward during both the expansion and contraction, and there are histories with equal probability in which time runs backward in the contracting phase. 

Not long after the Page response and later Laflamme work \cite{lafl}, Hawking abandoned the idea of time-symmetric histories \cite{laf} and considers only asymmetric histories that the quantum state contains. In this scenario individual histories would not experience a reversed arrow of time after the moment of the maximum expansion. However, the CPT invariance of the wave function of the universe ensures the presence of the time-reverse histories in the quantum states. Simply say, , if the quantum state of the universe contains an individual history than it must also contain with the same probability its time-reversed counterpart. This is the result we should have in mine in what follows.

\section{ The conditional probability interpretation}

Prior to delve into examining the relevance of time-reversed histories that are contained in the quantum state of the universe we should briefly describe the main concepts of conditional probability interpretation of time that will be useful for the rest of the paper.

We start recalling that the PaW approach \cite{pag}, \cite{woo} stems from the necessity to overcome the problem of time \cite{ish} , \cite{ande} that arises in canonical quantization of general relativity. The earliest days of canonical quantization of general relativity came with a big surprise. The Wheeler – De Witt equation (WDW) \cite{dew}, i.e. the quintessence of canonical quantization, fails to specify an external time for the dynamics of the universe. In contrast to our intuition about the evolving world around us, the Wheeler – De Witt equation,

\begin{equation}
H\ket{\Psi} = 0,
\end{equation} 

predicts that the physical state of the universe $\ket{\Psi}$, with the total Hamiltonian $H$, do not evolve in time. Surprisingly, the WDW equation (Eq.(1)) describes a static universe. This profound dilemma stimulates research in order to answer the crucial question: How to understand the evolution of the quantum state of the universe in the absence of the external parameter time? This fundamental question evolved along the years in what is now called “the problem of time”.

One particularly promising approach to overcome the canonical quantization problem of time is the PaW mechanism. In this approach the global static state of the Universe is considered as a bipartite quantum system composed from a clock system ($C$) and the rest of the universe system, named in what follows the system ($S$). The corresponding Hilbert spaces are $\mathcal{H_S}$ for the rest of the universe and $\mathcal{H_C}$ for the clock, respectively. Thus, the Hilbert space of the universe $\mathcal{H}$ is the tensor product, $\mathcal{H_C}\otimes \mathcal{H_S}$. We can remark here that time is operationally defined as a measurement of time operator $T$ on the clock Hilbert space, $\mathcal{H_C}$, such that $T\ket{t}=t\ket{t}$, with the corresponding eigenstatates $\ket{t}$ of the time operator and the associated eigenvalues $t$ as time indicated by the clock.

The timeless formalism of Page and Wootters assumes the universe is a pure state $\ket{\Psi}$  with the total Hamiltonian $H=H_C \otimes 1_S+1_C \otimes H_S$, where $H_C$ is the Hamiltonian of the clock and  $H_S$ the Hamiltonian of the system, respectively. One notable particularity when consider the operational time is that the clock have to behave perfectly \cite{gio}, \cite{mar}, \cite{bry}, \cite{smi}. The consequences of this statement are as follows. First, the clock states are perfectly distinguishable such that:

\begin{equation}
\bra{t_m}\ket{t_n} = \delta(t_m-t_n).
\end{equation}

Second, the perfect behavior of the clock imply that the clock can only run forward in time such that there is no chance that the clock can run backwards in time. To clarify this aspect of a clock that behaves perfectly let us follow the work of Unruh and Wald \cite{unr}. Accordingly, we should break up the spectrum of the operator $T$ into intervals of finite size, such that we have an infinite sequence of states $\ket{t_0},\ket{t_1},\ket{t_2},...$ Thus, for all $t>0$, and $m>n$ to say that the clock runs forward requires a nonvanishing amplitude to go from $\ket{t_n}$ to $\ket{t_m}$ in time $t$,

\begin{equation}
\bra{t_m} e^{-iH_Ct} \ket{t_n} \neq 0,
\end{equation}

and on the same time, a zero probability to go backward in time from $\ket{t_m}$ to $\ket{t_n}$,

\begin{equation}
\bra{t_n} e^{-iH_Ct} \ket{t_m} = 0.
\end{equation}

Moreover, in the case of a perfect clock, the time operator $T$ is canonically conjugate to the clock Hamiltonian, $H_C$,

\begin{equation}
[T,H_C] = i\hbar.
\end{equation}

This commutation relation generates translations of $T$ in the position of the clock, which with the help of Eq. (3) results in $\ket{t_m}=e^{-iH_Ct} \ket{t_n}$  or, equivalently, the usual Schrodinger equation

\begin{equation}
 i\hbar\frac{d}{dt}\ket{t}= H_C\ket{t}.
\end{equation}

In this scenario, the state of the rest of the Universe system, $\ket{\Psi_S(t)}$   at time $t$ as a conditional state of the global static system universe on the clock being in the state $\ket{t}$ is:

\begin{equation}
 \ket{\Psi_S(t)}= \bra{t}\ket{\Psi},
\end{equation}

where $\ket{\Psi_S(t)} \in \mathcal{H_S}$.

It is easy to verify that such conditioned state obeys the Schrodinger equation \cite{pag}: 

\begin{equation}
 i\hbar\frac{d}{dt}\ket{\Psi_S(t)}= H_S\ket{\Psi_S(t)}.
\end{equation}

Now, the history of the system $S$ (the rest of the universe) can be traced back through the entangled global stationary state:

\begin{equation}
 \ket{\Psi}= \int dt \ket{t}\ket{\Psi_S(t)},
\end{equation}

which is to say that within the PaW mechanism, the global stationary state $\ket{\Psi}$ includes the time history of system $S$ with respect to the clock system $C$. In this respect, an individual history of the rest of the universe is encoded in a sequence of states  $\ket{\Psi_S(t_0)}$, $\ket{\Psi_S(t_1)}$,…,$\ket{\Psi_S(t_i)}$ with a time ordering sequence of events $t_0<t_1<...<t_i$.

\section{ Time-reversed histories }

Let us recall here that the timeless state $\ket{\Psi}$  must contain, with the same probability, the time-revered history of the rest of the universe with respect to system clock $C$, as Page pointed out in \cite{page}. Thus, we would like to question at this point the meaning of a time-reversed history in the PaW mechanism framework. In other words, we like to ensure that time reversal is consistent with the timeless approach. Consequently, we will consider the time-reversal of an individual history of the rest of the universe, i.e. an individual history of the system $S$ for which the eigenstates of system clock are measured in reverse order. 

We emphasized that according to Eq. (3) and Eq.(4) a perfect clock runs forward with no chance of a reversion of the arrow of time. Thus, how can we consider time-reversal of an individual history in this situation? Nevertheless, to resolve this issue we relate our discussion once again to the Unruh and Wald paper \cite{unr}. Accordingly, they concluded that the nonzero probability of the clock to run forward in time is related to the vanishing amplitude of the clock to go backward in time. We can easily see this by considering the amplitude to go from $\ket{t_n}$ to $\ket{t_m}$ in time $t$ as:

\begin{equation}
\bra{t_m} e^{-iH_Ct} \ket{t_n}={\bra{t_n} e^{iH_C\tilde{t}} \ket{t_m}}^*={\bra{{t_n}^*} e^{-iH_C{t}^*} \ket{{t_m}^*}}=0,
\end{equation}

for all $t>0$ ,$m>n$ and with $t^*=-t$. 
Unruh and Wald interpreted this relation in terms of a realistic clock that runs forward in time which must have a nonvanishing probability to run backward in time. 
In the line with our exposition we would reinterpret the relation (10) accordingly and we consider it as a consequence of the fact that the quantum state must include the time reversed history. Such that, we have a perfect clock that runs forward but, with the same probability we must have also, a perfect clock that runs backward in time. 

If we question the distinguishability of the basis states we see that,

\begin{equation}
\bra{t_m}\ket{t_n} = {\bra{t_n}\ket{t_m}}^* = \bra{{t_n}^*}\ket{{t_m}^*}=\delta(t_m-t_n),
\end{equation}

is verified for all ${t}^*=-t$ since ${\bra{{t_n}^*}\ket{{t_m}^*}}=\delta({t_n}^*-{t_m}^*)$. Accordingly, the assumption that the hands of the clock are running backward in time does not change the condition of orthogonality of the clock states. 

With all this in mind, let us now consider the situation of a clock that runs backward in time without any chance that the clock will run forward in time. In this scenario, the translations of $T$ when the hands of the clock run in reverse, taking into account relation (10) is: 

\begin{equation}
\ket{{t_n}^*}=e^{iH_Ct}\ket{{t_m}^*},
\end{equation}

which is a solution of time-reversed Schrodinger equation for all ${t}^*=-t$, 

\begin{equation}
 -i\hbar\frac{d}{dt}\ket{{t}^*}= H_C\ket{{t}^*}.
\end{equation}

From Eq.(12) we can trace back the evolution of the clock system $C$ as a time-reversed ordering sequence ${t^*}_i<...<{t^*}_1<{t^*}_0$ of the eigenvalues of the clock system. Moreover, since $t^*=-t$, we have the backward time ordering sequence ${-t}_i<...<{-t}_1<{-t}_0$. 

The obvious consequence of this backward time evolution is that it should affect the state of the system $S$ since the state of the rest of the universe system consist in conditioning a state of the universe on the clock states. We have seen that a particular history of the rest of the universe is encoded in a sequence of states that is related to the time ordering $t_0<t_1<...t_i$. In the case of the time-reversed ordering sequence ${t^*}_i<...<{t^*}_1<{t^*}_0$ we should expect a time reversed history of the system $S$ , with states of the form $\ket{{\Psi_S}^*({t}^*)}$ obtained by projecting $\ket{\Psi}$ onto $\ket{t^*}$. In other words, we expect the relative state of the rest of the universe to satisfy a time-reversed Schrodinger equation.

\section{ The time-reversed Schrodinger Equation}

We recall that the time-reversed history of the system $S$ is also contained with the same probability, in the state $\ket{\Psi}$. That is, the state of the rest of the universe at time $t^*$ which consist on conditioning a solution of $\ket{\Psi}$ on the clock being in state $\ket{{t}^*}$ now reads,  

\begin{equation}
 \ket{{\Psi_S}^*({t}^*)}= \bra{{t}^*}\ket{\Psi}.
\end{equation}

Now, to verify how the condition of the clock following a backward time ordering affect the evolution of the system $S$ (the rest of the universe) we have to act with $-\frac{d}{dt}$ on both sides of the Eq.(14) which yield:

\begin{equation}
\begin{aligned}
 \frac{d}{dt}\ket{{\Psi_S}^*({t}^*)} &=  \frac{d}{dt}\bra{{t}^*}\ket{\Psi}\\
 &=(\frac{d}{dt}\bra{{t}^*})\ket{\Psi}+\bra{{t}^*}(\frac{d}{dt}\ket{\Psi})\\
 &=(\frac{d}{dt}\bra{{t}^*})\ket{\Psi}\\
 &=-\frac{i}{\hbar}\bra{{t}^*} H_C \ket{\Psi}.
 \end{aligned}
 \end{equation}

We know that Hamiltonian of the system clock is  $H_C=H-H_S$ and since $H\ket{\Psi}=0$ we have further:

\begin{equation}
\begin{aligned}
 \frac{d}{dt}\ket{{\Psi_S}^*({t}^*)}   
 &=-\frac{i}{\hbar}\bra{{t}^*} H-H_S \ket{\Psi}\\
 &=\frac{i}{\hbar}H_S\bra{{t}^*}\ket{\Psi}\\
 &=\frac{i}{\hbar}H_S\ket{{\Psi_S}^*({t}^*)}.
 \end{aligned}
 \end{equation}

It can be noted by rearranging the terms in Eq.(16) that the relative state of the rest of the universe $\ket{{t}^*}$ satisfies the time-reversal Schrodinger equation 

\begin{equation}
 -i\hbar\frac{d}{dt}\ket{{\Psi_S}^*({t}^*)}= H_S\ket{{\Psi_S}^*({t}^*)},
\end{equation}
for ${t}^*=-t$.

Thus, we demonstrate that considering a backward evolution of the clock system is consistent with the rest of the universe system evolving according to a time-reversed Schrodinger equation, which fulfills our expectation. 

The time-reversed history of the system $S$ can be traced back from the entangled state 

\begin{equation}
 \ket{\Psi}= \int dt \ket{{t}^*}\ket{{\Psi_S}^*({t}^*)},
\end{equation}

as a sequence of reversed order events encoded in the states $\ket{{\Psi_S}^*({t_i}^*)}$,..., $\ket{{\Psi_S}^*({t_1}^*)}$,$\ket{{\Psi_S}^*({t_0}^*)}$ with a time ordering sequence $-t_i<...<-t_1<-t_0$.
At this point we can safely say that a particular history of the universe in which the system clock runs backward is consistent with the PaW mechanism. 

\section{ The emergent nature of probabilities}

An individual history of the system $S$, i.e. the rest of the universe, is the sequence of events $\ket{\Psi_S(t_0)}$, $\ket{\Psi_S(t_1)}$,…,$\ket{\Psi_S(t_i)}$. Along this particular history, any particular state $\ket{\Psi_S(t_i)}$ of the rest of the universe is conditioned on the clock being in a specific state $\ket{t_i}$.  Thus, the quantities $\bra{t_i}\ket{\Psi}$ should correspond to probability amplitude distributions associated to the probability of finding  $\ket{\Psi_S(t)}$ on the system S when we measure $\ket{t}$ on the clock system $T$.

However, the global static state of the universe $\ket{\Psi}$  must contain with the same probability, the time-reversed history of the system $S$. So to speak, the quantum state must contain not only the time forward histories, but with the same probability, the backward time histories of the universe as well.  In an individual time-reversed history, the same states of the rest of the universe are attained in the reversed order. Accordingly, a particular time-reversed history of system $S$, consists in the sequence of states $\ket{{\Psi_S}^*({t_i}^*)}$,...,$\ket{{\Psi_S}^*({t_1}^*)}$,$\ket{{\Psi_S}^*({t_0}^*)}$ , i.e. the reversed order of the same events contained in the time forward history. Over the time-reversed history of the system $S$ any particular state $\ket{{\Psi_S}^*({t_i}^*)}$  arises as a  conditional state of the universe on the clock being in the state $\ket{t_i^*}$ as measured on the clock system.  

Since now we have considered the set of histories and their time-reversed counterparts individually. We would like to probe the consequences of the fact that the global static state contains, with the same probability both set of histories. An obvious immediate result is that conditioning a solution on the global static state of the universe $\ket{\Psi}$ on a specific  state of the clock $\ket{t_i}$ leads to the state $\ket{\Psi_S(t_i)}$ related to the time forward history and also the state $\ket{{\Psi_S}^*({t_i}^*)}$ that is contained in the time backward history.  Accordingly, a conditional state of the universe when a specific time $t$ is measured on the clock system can be written as: 

\begin{equation}
 \bra{t}\ket{\Psi}\bra{{t}^*}\ket{\Psi}= \ket{\Psi_S(t)}\ket{{\Psi_S}^*({t}^*)}.
\end{equation}

At this point we should make some remarks over the Eq.(19). We can easily recognize on the left side the presence of probability amplitude to find $\ket{\Psi_S(t)}$ on the system $S$ at a specific time $t$ times the probability amplitude  to find the state $\ket{{\Psi_S}^*({t_i}^*)}$  on the same system $S$ at the same specific time $t$. This is an indication that the left hand side of Eq.(19) reflects the probability, such that we can write the left hand side of Eq.(19) as:

\begin{equation}
 \bra{t}\ket{\Psi}\bra{{t}^*}\ket{\Psi}= Pr_{\ket{\Psi_S(t)}},
\end{equation}

where $Pr_{\ket{\Psi_S(t)}}$ is the probability of finding the system $S$ in a particular state when the state of the universe is conditioned on finding time $t$ as indicated by the hands of the clock. 
In other words, a conditional state of the universe when a specific time $t$ is measured on the clock system reflects the probability of finding the rest of the universe in a specific state labeled by the time $t$, such that we have:

\begin{equation}
 Pr_{\ket{\Psi_S(t)}}= \ket{\Psi_S(t)}\ket{{\Psi_S}^*({t}^*)},
\end{equation}

which is consistent with the Born rule. 
We can draw here a remarkable conclusion. The probability nature of quantum world naturally emerges in the scenario when we consider, in the PaW mechanism approach that the quantum state of the universe contains both the time-forward and the time-reversed set of histories.   

As a final observation let us note here another important characteristic of our investigation. Indeed, once we assume the probability interpretation of the conditional state of the universe, the right hand side of the Eq.(19) ensures that:

\begin{equation}
 \ket{\Psi_S(t)}\ket{{\Psi_S}^*({t}^*)}=\bra{\Psi_S(t)}\ket{\Psi_S(t)}=1,
\end{equation}

which results in remarking that the system state $\ket{\Psi_S(t)}$ is properly normalized at all times.

\section{Conclusions}

To conclude, in canonical quantum gravity the quantum state of the universe contains with the same probability both a particular history and its time-reversal counterpart. We have investigated this statement in the context of conditional probability interpretation of a time. Accordingly, first we have shown that a clock system that runs backward in time is consistent with the Page and Wootters mechanism and we derived a time-reversed Schrodinger equation for the evolution of the rest of the universe system.

Secondly, considering that the same particular quantum state is acquired both in an individual history with time running forward and in its time-reversed history, we have shown that conditioning the state of the universe on the hands of the clock showing a specific time results in the probability of finding the rest of the universe system in a particular state. 

We were lead to the conclusion that probability nature of quantum world naturally emerges if we consider in the PaW mechanism approach that the quantum state of the universe contains both the time-forward and the time-reversed set of histories.

\end{document}